%% file: submission.tex
\def\fullMode{TRUE}
\ifx\fullMode\undefined \newcommand{\FullVer}[1]{}
\newcommand{\ConfVer}[1]{#1} \else \newcommand{\FullVer}[1]{#1}
\newcommand{\ConfVer}[1]{} \fi

\documentclass[10pt,a4paper]{article}
\usepackage{latexsym}
\usepackage[T1]{fontenc}
\usepackage[english]{babel}
\usepackage[nohead,includefoot,margin=2cm]{geometry}
\usepackage[compact,raggedright,small]{titlesec}
\usepackage[affil-it]{authblk}
\usepackage[runin]{abstract}
\usepackage{multicol}
\usepackage{blindtext}
\usepackage{setspace}
\usepackage{amsmath}
\usepackage{amssymb}
\usepackage{comment}
\usepackage[affil-it]{authblk}
\usepackage{graphicx}
\usepackage{caption}
\usepackage{subcaption}
\usepackage{hhline}
\usepackage{multirow}
\usepackage[stable]{footmisc}
\usepackage{verbatim}
\usepackage[T1]{fontenc}
\usepackage[utf8]{inputenc}
\usepackage{tikz}
\usetikzlibrary{arrows}

\addto{\Affilfont}{\small}

\usepackage{array}
\newcolumntype{P}[1]{>{\centering\arraybackslash}p{#1}}

\newcommand*\samethanks[1][\value{footnote}]{\footnotemark[#1]}

\newtheorem{theorem}{Theorem}[section]

\newtheorem{example}[theorem]{Example}

\title{Representation Independent Analytics Over Structured Data}
\date{\small (Dated August 20, 2014)}
\author[1]{Yodsawalai Chodpathumwan {\it ychodpa@illinois.edu} \thanks{These authors contributed equally to this work.}}
\author[2]{Jose Picado {\it picadolj@oregonstate.edu} \samethanks}
\author[2]{Arash Termehchy {\it termehca@oregonstate.edu}}
\author[2]{Alan Fern {\it afern@oregonstate.edu}}
\author[3]{Yizhou Sun {\it yzsun@ccs.neu.edu}}

\affil[1]{School of EECS, Oregon State University, Corvallis, OR, USA}
\affil[2]{College of Computer and Information Science, Northeastern University, Boston, MA, USA}
\affil[3]{Department of Computer Science, University of Illinois, Urbana, IL, USA}

\begin{document}

\maketitle

\maketitle

\begin{abstract}
Database analytics algorithms leverage quantifiable structural 
properties of the data
to predict interesting concepts and relationships. 
The same information, however, can be represented using many 
different structures and the structural properties observed over particular representations do not necessarily hold for alternative  
structures. Thus, there is no guarantee that 
current database analytics algorithms will still provide 
the correct insights, no matter what structures 
are chosen to organize the database.
Because these algorithms tend to 
be highly effective over some choices of 
structure, such as that of the databases 
used to validate them, but not so effective with others, 
database analytics has largely remained the province of experts
who can find the desired forms for these algorithms. 
We argue that in order to make database analytics 
usable, we should use or develop
algorithms that are effective over a 
wide range of choices of structural organizations. 
We introduce the notion of {\it representation independence}, 
study its fundamental properties for a wide range of 
data analytics algorithms, and empirically analyze the amount of 
representation independence of some popular database analytics
algorithms. Our results indicate that most algorithms 
are not generally representation independent 
and find the characteristics of more 
representation independent heuristics 
under certain representational shifts. 

\end{abstract}
\input{intro.tex}

\input{related.tex}
\input{framework.tex}

\input{experiments.tex}

\section{Conclusion and Future Work} 
\label{section:conclusion}
Since current DB analytics algorithms 
generally rely on representational details of DBs, 
they are effective only over certain 
representations of the data.
We argued that a usable algorithm should 
deliver the same results no matter
which representation is chosen for the data and studied the 
fundamental properties of this notion. Our empirical studies
provided new insights on the characteristics of  
some robust DB analytics algorithms.
We plan to develop more representation independent algorithms for the discussed problems. 

{\
\bibliographystyle{abbrv}
\bibliography{ref}
}
\end{document}

%% file: intro.tex
\section{Introduction}
\label{section:introduction}

Over the last 20 years, users' information needs over structured data
expanded from
seeking exact answers to precise queries
to finding entities or patterns {\it similar}
to a given entity or pattern, discovering {\it interesting}
entities and patterns, or predicting {\it novel}
relations and concepts \cite{Hellerstein:2012:PVLDB,Han:DataMining,Getoor:SRLBook,Jeh:KDD:02,Tong:ICDM:06}.
As part of its response, the research community proposed a multitude of
supervised and unsupervised algorithms to solve exploration
and analytics problems over structured data in different contexts,
such as similarity query processing, inexact pattern matching,
and concepts and relationship prediction
\cite{Jeh:KDD:02,Tong:ICDM:06,Getoor:SRLBook,Han:DataMining}.
Since the properties of interesting and desirable answers are no
longer precisely defined in the query, and in many cases there
is no query at all in the traditional sense,
these algorithms use intuitively appealing heuristics
to choose, from among all possible answers,
those that are most interesting, important, and likely
to satisfy the user's information need.

The power of database 
exploration and analytics remains out of the
reach of most users, however, as today's
database exploration and analytics algorithms and tools are usable
only by highly trained database analysts who can
predict which algorithms are likely to be effective for
particular representations of the data, 
or who are able to customize these algorithms to satisfy 
their information needs over a new database.
To see why, consider the following examples.


\begin{figure*}[ht]
\begin{minipage}[b]{0.4\textwidth}
\centering
\includegraphics[width = \columnwidth]{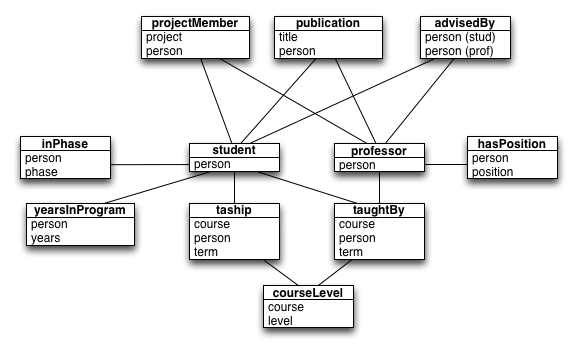}
\end{minipage}
\begin{minipage}[b]{0.4\textwidth}
\centering
\includegraphics[width=\columnwidth]{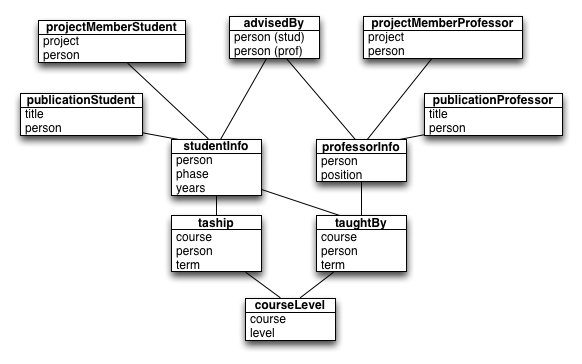}
\end{minipage}
\begin{minipage}[b]{0.19\textwidth}
\centering
\includegraphics[width=\columnwidth]{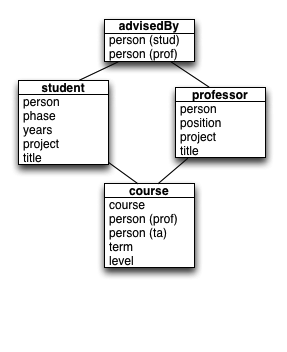}
\end{minipage}
\caption{\small Various Schemas for the UW-CSE database.}
\label{figure:uwcse_schemas}
\end{figure*}

\noindent
\begin{example}
{\bf Relational Learning:} Given a database and 
training instances of a new target relation, 
relational learning algorithms attempt to induce general (approximate) definitions of the 
target in terms of existing relations.  
For example, given a database with student and professor information, the goal may be to 
induce a Datalog definition of a missing relation {\it advisedBy(stud,prof)} based
on a training set of known student-advisor pairs. Since the space of possible definitions 
(e.g. all Datalog programs) is enormous, learning algorithms employ heuristics to search 
for effective definitions, which generally depend on the schema 
of the database. More generally, statistical relational learning 
algorithms\cite{Getoor:SRLBook} use the same heuristic mechanisms for structure learning, which in turn renders them schema dependent. 

As an example, Figure~\ref{figure:uwcse_schemas} shows the relations from three schemas for the 
UW-CSE\footnote{http://alchemy.cs.washington.edu/data/uw-cse}
database over students and professors, which is used as a common relational learning
benchmark. The first and original schema was designed by relational learning 
experts and is (almost) in sixth
normal form, which is generally discouraged 
due to its poor usability and
performance in query processing \cite{AliceBook}.
A database designer may use a schema closer to 
the alternative schemas in Figure~\ref{figure:uwcse_schemas}. 
We used the popular learning algorithm FOIL~\cite{Quinlan:FOIL} to induce a definition of {\it advisedBy(stud,prof)}
for each of the two schemas, resulting in two very different definitions. The original schema yielded a  
more accurate definition based on co-authorship information, while the alternative schema led to a definition based 
on information about TA and teaching assignments.  
\end{example}
\begin{example}
{\bf Graph Analytics:} Figure~\ref{fig:FreebaseIMDB} (a) and (b) show excerpts of IMDb ({\it imdb.com}) and
Freebase ({\it freebase.com}) about the same set of 
movies, their characters, and the actors who played them, respectively. 
IMDb and Freebase use different representations to express the same
relationships between movies. Various link-based similarity search algorithms over graph databases, such as Random Walk with Restart (RWR) \cite{Tong:ICDM:06} and SimRank \cite{Jeh:KDD:02}, 
use proximity-based heuristics. RWR evaluates how likely an entity will be visited if a random surfer starts and keeps re-starting from the query entity. SimRank evaluates the similarity between two entities 
according to how likely two random surfers will meet each other if they start from the two entities.
RWR and SimRank 
find {\it Star Wars III} more similar to
{\it Star Wars V} than to  {\it Jumper}
in Figure~\ref{fig:FreebaseIMDB}(a), but
find {\it Star Wars III} to be more similar to {\it Jumper} in Figure~\ref{fig:FreebaseIMDB}(b).
\end{example}
\input{fig_schemaimdb-real.tex}

Generally, there is no 
canonical representation for a particular set of content
and people often represent the same information in
different structures~\cite{AliceBook,infopreserve:XML,infopreserve:hull}.
For example, many researchers use DBLP ({\it dblp.uni-trier.de}),
which is a publicly available bibliographical database about
publications in computer science with rather a few types of 
entities, to evaluate their algorithms. 
We have observed that different researchers often represent this dataset differently.
Fig.~\ref{fig:schema-dblp} shows two different representations for
DBLP in two papers \cite{Sun:VLDB:11,Zhao:CIKM:09} published in premier database venues.
\input{fig_schemadblp-orig.tex}

Thus, users generally have to restructure their databases
to some {\it proper representations}, in order to
effectively use database analytics algorithms,
i.e., deliver
the insights that a domain expert would judge as relevant and
important. To make matters worse, these algorithms do not normally
offer any clear description of their desired representations and
database analysts have to rely on their own
expertise and/or do trial and error to find such representations.
Nevertheless, we want our database analytics algorithms
to be used by ordinary users, not just experts.
Further, the structure
of large-scale databases constantly evolve
\cite{Cohen:2009:MSN:1687553.1687576},
and we want to move away from the need for
constant expert attention to keep exploration algorithms effective.

For similar reasons, current database analytics algorithms will
not be well suited for Big Data, which is inherently heterogeneous
and evolving as it obtains its content from many different data sources. Moreover, researchers often use 
analytics algorithms, such (statistical) relational learning and mining techniques, to solve
various important data management problems, 
such as entity resolution, 
information extraction, and query processing \cite{Abouzied:2013:PODS,Getoor:SRLBook}.
Thus, the issue of representation dependence
appears in other areas of data management.

To cope with organizational heterogeneity and evolution in
large-scale data, it is time to move beyond database 
analytics algorithms that are effective only over certain 
representations of the database.
To this end, we propose a novel approach to
database analytics that considers {\it representation independence},
i.e., the ability to deliver the same answers regardless
of the choices of structure for organizing the data.
{\bf We argue for considering the robustness of an algorithm 
across various representations 
to better understand its usability.}
Further, considering the authors' recent success in developing
effective keyword search techniques that are robust across
multiple representations of the same information \cite{SchemaIndep:TKDE},
{\bf we believe
that it is possible to provide ordinary
users with an 
arsenal of effective database analytics methods that are robust
across multiple representations of the same information.}
In particular:
\begin{itemize}
\item
We introduce a novel framework that
defines, quantifies, and analyzes the amount of representation 
independence of database analytics algorithms. 

\item
We empirically study the 
representation independence of some popular
relational learning and graph analytics algorithms.
Our investigations indicates that 
most these algorithm largely depend on representational
details, while a couple of them have reasonable amount of 
representation independence.
It also offers promising insights for developing 
more representation independent algorithms.
\end{itemize}

This paper is organized as follows.
Section~\ref{section:related} describes the related work.
Section~\ref{section:framework} introduces our proposal
of representation independent data analytics.
Section~\ref{section:empirical} contains our preliminary results on
the amount of representation independent of some well known
data analytics algorithms and
Section~\ref{section:conclusion}
concludes the paper.

%% file: fig_schemaimdb-real.tex
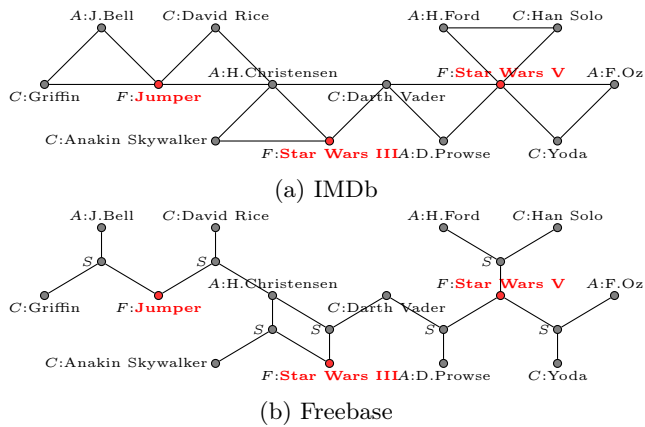
\begin{figure}[h!]
\centering
\begin{subfigure}[b]{1.00\textwidth}
\centering
\begin{tikzpicture}[->,>=stealth',scale=0.75]
\tikzstyle{every node}=[circle,draw,fill=black!50,inner sep=1pt,minimum width=3pt]
\tikzstyle{every label}=[rectangle,draw=none,fill=none,font=\tiny]
\node (GF) [label=below:$C$:Griffin] at (-2,0) {};
\node[fill=red!80] (JP) [label=below:{\tiny $F$}:\textcolor{red}{\bf Jumper}] at (0,0) {};
\node (HC) [label=above:$A$:H.Christensen] at (2,0) {};
\node (DV) [label=below:$C$:Darth Vader] at (4,0) {};
\node[fill=red!80] (W5) [label=above:{\tiny $F$}:\textcolor{red}{\bf Star Wars V}] at (6,0) {};
\node (FO) [label=above:$A$:F.Oz] at (8,0) {};
\node (JB) [label=above:$A$:J.Bell] at (-1,1) {};
\node (DR) [label=above:$C$:David Rice] at (1,1) {};
\node (HF) [label=above:$A$:H.Ford] at (5,1) {};
\node (HS) [label=above:$C$:Han Solo] at (7,1) {};
\node (AS) [label=left:$C$:Anakin Skywalker] at (1,-1) {};
\node[fill=red!80] (W3) [label=below:{\tiny $F$}:\textcolor{red}{\bf Star Wars III}] at (3,-1) {};
\node (DP) [label=below:$A$:D.Prowse] at (5,-1) {};
\node (LF) [label=below:$C$:Yoda] at (7,-1) {};
\draw [-] (JP) -- (HC) -- (DV) -- (W5) -- (FO) -- (LF) -- (W5);
\draw [-] (JP) -- (DR) -- (HC) -- (W3);
\draw [-] (W5) -- (HF) -- (HS) -- (W5);
\draw [-] (HC) -- (AS) -- (W3) -- (DV) -- (DP) -- (W5);
\draw [-] (JP) -- (GF) -- (JB) -- (JP);
\end{tikzpicture}
\caption{IMDb}
\label{fig:imdb}
\end{subfigure}
\begin{subfigure}[b]{1.00\textwidth}
\centering
\begin{tikzpicture}[->,>=stealth',scale=0.75]
\tikzstyle{every node}=[circle,draw,fill=black!50,inner sep=1pt,minimum width=3pt]
\tikzstyle{every label}=[rectangle,draw=none,fill=none,font=\tiny]
\node (GF) [label=below:$C$:Griffin] at (-2,0) {};
\node[fill=red!80] (JP) [label=below:{\tiny $F$}:\textcolor{red}{\bf Jumper}] at (0,0) {};
\node (HC) [label=above:$A$:H.Christensen] at (2,0) {};
\node (DV) [label=below:$C$:Darth Vader] at (4,0) {};
\node[fill=red!80] (W5) [label=above:{\tiny $F$}:\textcolor{red}{\bf Star Wars V}] at (6,0) {};
\node (FO) [label=above:$A$:F.Oz] at (8,0) {};
\node (JB) [label=above:$A$:J.Bell] at (-1,1.2) {};
\node (DR) [label=above:$C$:David Rice] at (1,1.2) {};
\node (HF) [label=above:$A$:H.Ford] at (5,1.2) {};
\node (HS) [label=above:$C$:Han Solo] at (7,1.2) {};
\node (AS) [label=left:$C$:Anakin Skywalker] at (1,-1.2) {};
\node[fill=red!80] (W3) [label=below:{\tiny $F$}:\textcolor{red}{\bf Star Wars III}] at (3,-1.2) {};
\node (DP) [label=below:$A$:D.Prowse] at (5,-1.2) {};
\node (LF) [label=below:$C$:Yoda] at (7,-1.2) {};
\node (s1) [label=left:$S$] at (1,0.6) {};
\node (s2) [label=left:$S$] at (2,-0.6) {};
\node (s3) [label=left:$S$] at (3,-0.6) {};
\node (s4) [label=left:$S$] at (5,-0.6) {};
\node (s5) [label=left:$S$] at (6,0.6) {};
\node (s6) [label=left:$S$] at (-1,0.6) {};
\node (s7) [label=left:$S$] at (7,-0.6) {};
\draw [-] (s1) -- (JP);
\draw [-] (s1) -- (DR);
\draw [-] (s1) -- (HC);
\draw [-] (s2) -- (HC);
\draw [-] (s2) -- (AS);
\draw [-] (s2) -- (W3);
\draw [-] (s3) -- (W3);
\draw [-] (s3) -- (DV);
\draw [-] (s3) -- (HC);
\draw [-] (s4) -- (DV);
\draw [-] (s4) -- (W5);
\draw [-] (s4) -- (DP);
\draw [-] (s5) -- (W5);
\draw [-] (s5) -- (HF);
\draw [-] (s5) -- (HS);
\draw [-] (s6) -- (GF);
\draw [-] (s6) -- (JP);
\draw [-] (s6) -- (JB);
\draw [-] (s7) -- (W5);
\draw [-] (s7) -- (FO);
\draw [-] (s7) -- (LF);
\end{tikzpicture}
\caption{Freebase}
\label{fig:freebase}
\end{subfigure}
\caption{{\small Fragments of IMDb and Freebase, where $A$, $C$, $F$, and $S$ refer to $actor$, $character$, $film$ and $starring$, respectively.}}
\label{fig:FreebaseIMDB}
\end{figure}

%% file: fig_schemadblp-orig.tex
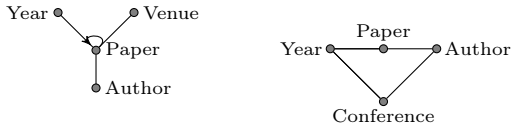
\begin{figure}[h!]
\centering
\begin{subfigure}[b]{0.22\textwidth}
\centering
\begin{tikzpicture}[->,>=stealth',scale=0.5]
\tikzstyle{every node}=[circle,draw,fill=black!50,inner sep=1pt,minimum width=3pt,font=\scriptsize]
\node (paper) [label=right:Paper] at (0,0) {};
\draw [-] (paper) -- + (-1,1)	node (term) [label=left:Year] {};
\draw [-] (paper) -- + (1,1)	node (venue) [label=right:Venue] {};
\draw [-] (paper) -- + (0,-1) node (author) [label=right:Author] {};
\draw [-] (paper) edge [loop] (paper);
\end{tikzpicture}
\end{subfigure}
\begin{subfigure}[b]{0.22\textwidth}
\centering
\begin{tikzpicture}[->,>=stealth',scale=0.7]
\tikzstyle{every node}=[circle,draw,fill=black!50,inner sep=1pt,minimum width=3pt,font=\scriptsize]
\tikzstyle{every label}=[rectangle,draw=none,fill=none,font=\scriptsize]
\node (paper) [label=above:Paper] at (1,0) {};
\node (author) [label=right:Author] at (2,0) {};
\node (conference) [label=below:Conference] at (1,-1) {};
\node (year) [label=left:Year] at (0,0) {};
\draw [-] (paper) -- (author);
\draw [-] (author) -- (conference);
\draw [-] (conference) -- (year);
\draw [-] (year) -- (conference);
\draw [-] (year) -- (paper);
\draw [-] (paper) -- (year);
\end{tikzpicture}
\end{subfigure}
\caption{{\small Different schemas for DBLP from \cite{Sun:VLDB:11} and \cite{Zhao:CIKM:09}.}}
\label{fig:schema-dblp}
\end{figure}

%% file: related.tex
\section{Related Work}
\label{section:related}

Researchers have proposed declarative 
languages to achieve physical data independence 
for data analytics so developers do not 
worry about how the data is 
physically stored and what the execution environment is  
\cite{Hellerstein:2012:PVLDB,DBLP:journals/debu/BorkarBCRPCWR12,Ghoting:2011:SDM:2004686.2005625}. 
We, however, focus on the independence from 
logical representation.

The architects of relational model have argued for
logical data independence, which  
oversimplifying a bit, means that an exact query should 
return the same answers no matter which logical schema is 
chosen for the data \cite{AliceBook,Codd:Computerworld:85}.
The idea of representation independence
extends the principle of logical data independence
for database (DB) analytics algorithms. These ideas also
differ in a couple of subtle but important issues.
In our approach, the DB does not need to have a 
predefined schema, as many DB analytics algorithms 
explore DBs without any well-defined schema.
One may achieve logical data independence 
by an affordable amount of experts' intervention, such 
as defining a set of views over the DB \cite{AliceBook}. 
However, it generally takes more time and much deeper
expertise to find the proper representation for 
a data analytics algorithm. Particularly, 
DB analytics applications normally contain more than a 
single algorithm \cite{Hellerstein:2012:PVLDB}, therefore, 
it will require a group of experts 
to achieve representation independence in this fashion. 
Hence, it is less likely to 
achieve representation independence via 
an affordable degree of 
expert's intervention.

Researchers have studied the issue of representational 
invariance in the context of probabilistic inference \cite{Salmon:Vindicating:63,Halpern:JMLR:04}.
They measure the invariance of a probabilistic 
inference method as its robustness against certain
representational transformation(s). We, however, study this issue
for supervised and unsupervised mining and learning algorithms. 
Researchers in this line of work measure
the representation independence of 
a method under certain transformation as
a binary value. In this paper, we propose non-binary 
metrics for representation independence, which make it easier
to compare the robustness of various algorithms 
under same type of representational shifts. 
We also focus on representational variations 
and analytics algorithms over structured data models. 

Researchers have analyzed the stability of 
some DB analytics algorithms 
against relatively small perturbations 
in the data \cite{Ng2001SAL,Getoor:SRLBook}.
We also seek to instill 
robustness in DB analytics algorithms, 
but we are targeting robustness in a new dimension:  
robustness in the face of variations 
in the representation of data. 

Finding a subset of relevant 
features from the data, 
is an important step in deploying learning algorithms \cite{Anderson:CIDR:2013}.
When applied in this context, 
the idea of representation independence prefers 
features that are more robust against 
representational variations in the underlying DB. 

We reported some initial results on the representation 
independence of similarity search algorithms over graphs in 
\cite{Chopathumwan:2014:GRADES}. Current paper extends
the idea of representation independence for general
DB analytics algorithms and 
other data models, e.g. relational model, and 
provides the required formal framework for them.
It compares other properties of algorithms, such as their running times, across various representations. 
It studies DB analytics algorithms for other problems, 
such as relational learning.
We also analyze the amount of representation independence
for graph analytics algorithms under some new  
representational shifts and DBs 
and find some algorithms to be  
representation independent under them, as opposed to 
the empirical studies in \cite{Chopathumwan:2014:GRADES}.
Finally, we analyze the characteristics of the
representation independent algorithms for both graph analytics
and relational learning.


%% file: framework.tex
\section{Representation Independence}
\label{section:framework}

Intuitively, a representation independent 
algorithm should return the same answers 
across databases that represent the same information.
We define the conditions under 
which different databases represent the same information.
{\it Database transformations} have been used to 
model the relationship between different choices of structure 
in terms of the amount of information they can contain \cite{infopreserve:XML}.
More formally, \textbf{transformation} $\tau$ over DB $D$ 
is a (computable) 
function that maps $D$ to another DB $\tau(D)$.
If $\tau$ is {\it invertible}, $\tau(D)$ carries at least as 
much information as $D$. In other words, we can reconstruct the information available in $D$, given the transformed DB. 
For example, the transformation between IMDb and 
Freebase DBs in Figure~\ref{fig:FreebaseIMDB}
is invertible, as it replaces every {\it triangular} 
subgraph whose nodes represent entities of types {\it film}, 
{\it actor}, and {\it character} in a graph database with a {\it star} 
subgraph where these entities are connected
via a single instance of type {\it starring}. 
If there is also an invertible transformation 
$\sigma$ from $\tau(D)$ to $D$, $D$ and $\tau(D)$ 
represent the same information and are called {\it equivalent}. 
We can similarly define the notion of information 
equivalence between two schemas. 
Let $I(S)$ be the set of all DBs with schema $S$.
Given schemas $S_1$ and $S_2$,
if there an {\it invertible} transformation 
$\tau$: $I(S_1) \rightarrow I(S_2)$, and invertible
transformation $\sigma$: $I(S_2) \rightarrow I(S_1)$,
$S_1$ and $S_2$ are equivalent.
If $\tau$ and $\sigma$ belong to a family (set) of 
transformation $T$, e.g. vertical decompositions over 
relational schemas,
we call schemas $S_1$ and $S_2$, and their mapped 
DBs, equivalent under $T$.  
 
Many data analytics algorithms over structured data
assume that their answers belong to a 
fixed {\it hypothesis language}, for example
a relational learning algorithm may assume that 
its target concepts are Datalog expressions 
without recursion \cite{Getoor:SRLBook}.
In order for an algorithm to deliver the same 
answers over equivalent DBs
under a family of transformations $T$, 
$T$ must also establish a bijective 
mapping between the hypothesis space 
over different representations. 
One may extend the work on query 
preservation \cite{infopreserve:XML}, 
to check this property. 

Some DB analytics algorithms, such as (statistical) relational learning and graph analytic methods, 
treat the values stored in the database 
as uninterpreted atomic object \cite{Getoor:SRLBook,Sun:VLDB:11,Tong:ICDM:06,Tong:ICDM:06}.
Some invertible transformations 
may modify or perform some arithmetic operations 
on the values in the database.
It is not reasonable to expect the algorithms 
that treat values as atomic objects to return the same
answers over representational shifts that modify values.
Hence, we focus on the algorithms and 
transformations that treat values as atomic objects and 
leave other types of transformations
as future work.

Generally, algorithm $A$ is {\it representation independent} under 
family of transformations $T$ iff it returns the same answers 
over equivalent DBs under $T$ for the same input. 
The answers across equivalent DBs may 
follow different representations, but convey 
the same information.
The notion of input may be interpreted 
differently in different analytics problems
For instance, it is a query for some DB analytics algorithms,
such as finding similar entities, and 
a set of training data
for supervised tasks, such as learning a novel relation. 

Provided that algorithms $A$ and $B$ are representation 
independent under families of transformations $T_A$ and $T_B$ respectively, $A$ is {\it more representation independent} than 
$B$ if  $T_A \supset T_B$. In other words,  
more representation independent algorithms are
robust under a wider range of representational changes.
For instance, a relational learning algorithm 
that is robust under both vertical and horizontal decompositions
is more representation independent than the one that is robust 
only under vertical decomposition.
Since perfect representation independence 
under a certain family of transformations
may be hard to achieve
for some analytic tasks, one may like to find the 
{\it most representation independent}
algorithm for some analytic task under a certain set of transformations.
Given a family of transformations $T$, algorithm $A$ is
{\it more representation independent} than algorithm $B$
if $A$ returns {\it more similar} answers than $B$
across DBs that are equivalent under $T$.
The amount of similarity between answers may be computed
differently in different analytic tasks. For instance, 
the algorithms that find the most related entities to 
a given entity in a DB, return a ranked list of answers.
Thus, one may use a distance metric for ranked lists
to compute the 
similarities of the answers across equivalent DBs.

One may add additional constraints to the 
definition of representation independence
in addition to the equality or similarity of answers. 
Our empirical studies 
of many DB analytics algorithms indicate that
they need considerable amount of parameter (re-)tunning
across various representations. Ideally, 
we would like a representation independent 
algorithm to have almost the same configuration across
equivalent DBs, thus, a stronger type of 
representation independence is for an algorithm to have a 
minimal amount of reconfiguration across equivalent DBs.
We have also found some algorithms to be very inefficient
over certain representations, therefore, 
one may extend the definition of representation 
independence over family of transformations $T$ such that
the it returns the same or similar answers over
equivalent DBs under $T$ {\it within a reasonable amount of time}.

%% file: experiments.tex
\section{Empirical Study}
\label{section:empirical}

One approach to create 
representation independence is to 
run an algorithm and over all representations 
of a validation subset of the DB and select the 
representation with the most accurate answers. 
Nonetheless, computing all representations 
of a DB is generally undecidable \cite{infopreserve:XML}.
If confined within a particular family of 
transformations, DBs have normally 
an exponential number of representations, 
for example a relational table may have 
exponential number of distinct vertical decompositions.
This number will be larger for the DBs that do not
follow a fixed schema.
As many algorithms need some time for parameter tunning under 
a new representation \cite{MLBase:CIDR}, it may take a 
prohibitively long time to find the best representation.
Moreover, such a validation subset is not generally available for 
unsupervised methods.
This approach also requires that the underlying DB
be transformed to the desired representation, 
which may not be practical for a large and/or 
constantly evolving DB that is being used
by varieties of algorithms, where each is effective
over a different representation.

Another method is to 
define a {\it universal representation} to which all possible 
representations can be transformed 
and use/develop algorithms that are effective 
over this representation. 
The experience gained from the idea of universal relation, 
indicates that even for DBs with fixed schema  
such representation may not always exist \cite{AliceBook}. 
Users also have to transform their DB to the universal
representation, which may be quite complex considering the 
intricacies associated with defining such a representation
and not practical for a large DB.

Hence, a more sustainable approach is to develop algorithms
with reasonably high degree of representation independence.  
In this section, we analyze and compare the representation 
independence of popular relational learning and graph analytics
algorithms to find characteristics of more representation 
independent heuristics.

\subsection{Relational Learning}

\input{exp-relationallearning.tex}

\subsection{Graph Analytics}
\input{exp-graphmining.tex}

%% file: exp-relationallearning.tex
We evaluate the impact of representation on three popular relational learning algorithms: 
FOIL~\cite{Quinlan:FOIL}, Progol~\cite{progol} and ProGolem~\cite{progolem}. FOIL is a greedy algorithm that induces one Datalog rule at a time by greedily specializing rules via adding literals to their bodies. Progol is similar to FOIL, but performs a more complete, non-greedy search over rule bodies within a restricted hypothesis space. 
ProGolem follows an alternative specific-to-general approach that searches for rules by combining pairs of overly specific rules into more general rules. 
We emulate both FOIL and Progol using Aleph\footnote{http://www.cs.ox.ac.uk/activities/machlearn/Aleph/aleph.html}, a well known Inductive Logic Programming (ILP) system. 
ProGolem is implemented in GILPS\footnote{http://www.doc.ic.ac.uk/~jcs06/GILPS/}, another ILP system.

In Aleph, we use the following configuration: $noise = 100\%$, $minpos = 2$, $search = heuristic$, $evalfn = compression$, $nodes = 10000$, $clauselength = 5$, and $i = 2$. We use the default values for the rest of the parameters except for $openlist$ (beam), which we set to $1$ to emulate FOIL and $inf$ to emulate Progol.
In ProGolem, we use the default configuration, except for the following parameters: $noise = 100\%$ and $i = 1$. The number of layers of new variables (parameter $i$) in Aleph is one unit higher than in ProGolem. Hence the values $i = 2$ in Aleph and $i = 1$ in ProGolem.

We run experiments using the UW-CSE and IMDb DBs. 
The UW-CSE data set consists of 12 relations, 2673 tuples, and 113 positive examples. We ignore the relation $tempAdvisedBy$. 
For both DBs, we generate negative examples using the closed-world assumption, and then sample these to obtain twice as many negative examples as positive examples. 
We divide the UW-CSE data set into 5 folds, each one corresponding to a different group of the CSE department, and 
we learn the relation {\it advisedBy(stud,prof)},
which indicates that student {\it stud} is advised by professor {\it prof}. 
We represent the DB using three equivalent schemas: 
the original schema, which is almost in 6th normal form, 
its transformed 4th normal form, and its denormalized schema, shown in Figure~\ref{figure:uwcse_schemas}.
These schemas are mapped to each other via vertical decomposition (i.e. projection) and composition (i.e. join) transformations.

\begin{figure*}[htp]
\begin{minipage}[b]{0.5\textwidth}
\centering
\includegraphics[width = \columnwidth]{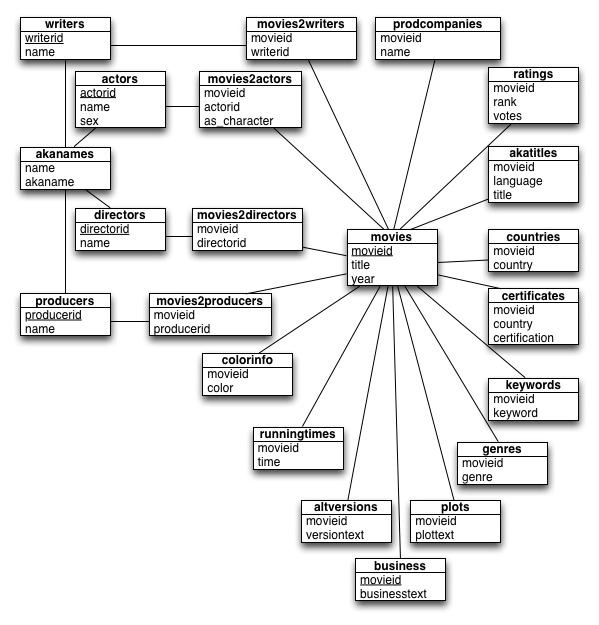}
\end{minipage}
\begin{minipage}[b]{0.5\textwidth}
\centering
\includegraphics[width=\columnwidth]{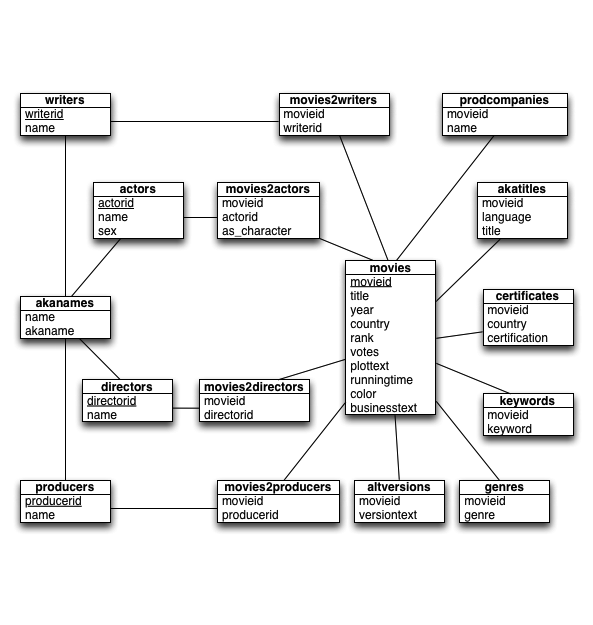}
\end{minipage}
\caption{\small Schemas for the IMDb DB.  }
\label{figure:imdb_schemas}
\end{figure*}

We download the relational DB for IMDb from JMDB\footnote{http://www.jmdb.de}, and obtain a subset of 
the DB to create 5 folds, where each fold 
contains information about 100 randomly selected movies.
We remove the information about countries and genders from the DB, and learn the relations {\it countries(mov, ctry)}, which indicates that {\it mov} was filmed in {\it ctry}, and {\it femaleActor(actor)}, which indicates that {\it actor} is a female.
We use two equivalent schemas for IMDb: 
its original schema, which is (almost) in 6NF, 
and an alternative schema that contains rather important attributes of a {\it movie} in a single
table. We have picked these attributes based on the information in the main Web page of a movie
in IMDb Website. One may prefer to use such a schema in order to generate and retrieve the 
main page of a movie fast. Since this schema is quite close to a 4th normal form, we denote it by 4NF in this section.
These schemas are shown in Figure~\ref{figure:imdb_schemas}.
Because it takes a long time to learn relations 
over more denormalized schemas of IMDb, we did not use them. 

We evaluate accuracy, precision, recall, and running time, showing the average over 5-fold cross validation. Accuracy is the proportion of true results in all examples, precision is the proportion of true positives against all the positive results, and recall is the proportion of true positives against all positive examples. 
We have also measured the robustness of these methods in terms
of the number of equivalent rules learned over 
different representations. We have found out that this metric may
not reasonably reflect the amount of robustness for an algorithm
as a slight variation in the rule definition will result in
robustness of zero. 
Table \ref{table:uwcse-advisedby} and \ref{table:imdb-countries-femaleActor} show the results for the UW-CSE and IMDb DBs, respectively. 

\begin{table}[tbp]
{\scriptsize
\centering
\begin{tabular} {  P{0.15\columnwidth} | P{0.15\columnwidth} | P{0.15\columnwidth} | P{0.15\columnwidth} | P{0.15\columnwidth}  }
Algorithm & Metric & 6NF & 4NF & Denorm. \\
\hline
\multirow{4}{*}{FOIL} & Accuracy & {\bf 0.72} & {\bf 0.72} & 0.64 \\
& Precision & 0.62 & {\bf 0.64} & 0.48 \\
& Recall & 0.70 & {\bf 0.71} & 0.41 \\
& Time (s) & {\bf 2.44} & 2.77 & 396.48 \\
\hline
\multirow{4}{*}{Progol} & Accuracy & 0.73 & {\bf 0.77} & 0.71 \\
& Precision & {\bf 0.62} & {\bf 0.62} & 0.59 \\
& Recall & 0.72 & {\bf 0.81} & 0.52 \\
& Time (s) & 51.06 & {\bf 43.93} & 226.39 \\
\hline
\multirow{4}{*}{ProGolem} & Accuracy& {\bf 0.77} & 0.74 & 0.75 \\
& Precision & 0.83 & {\bf 0.86} & 0.83 \\
& Recall & {\bf 0.44} & 0.32 & 0.36 \\
& Time (s) & 65.10 & {\bf 56.31} & 271.30 \\
\end{tabular}}
\caption{{\small Results of learning relations over UW-CSE.}}
\label{table:uwcse-advisedby}
\end{table}
\begin{table}[tbp]
{\scriptsize
\centering
\begin{tabular} {  P{0.2\columnwidth} | P{0.2\columnwidth} | P{0.2\columnwidth} | P{0.2\columnwidth}  }
Algorithm & Metric & 6NF & 4NF \\
\hline
\multirow{4}{*}{FOIL} & Accuracy & 0.69 & {\bf 0.71} \\
& Precision & 0.48 & {\bf 0.66} \\
& Recall & 0.13 & {\bf 0.20} \\
& Time (m) & 129.10 & {\bf 122.08} \\
\hline
\multirow{4}{*}{Progol} & Accuracy & 0.70 & {\bf 0.71} \\
& Precision & 0.59 & {\bf 0.65} \\
& Recall & 0.17 & {\bf 0.23} \\
& Time (m) & 1269.39 & {\bf 851.10} \\
\hline
\multirow{4}{*}{ProGolem} & Accuracy & {\bf 0.69} & {\bf 0.69} \\
& Precision & {\bf 0.87} & 0.86 \\
& Recall & 0.08 & {\bf 0.09} \\
& Time (m) & {\bf 16.22} & 24.46 \\
\end{tabular}}
\caption{{\small Average results of learning relations over IMDb.}}
\label{table:imdb-countries-femaleActor}
\end{table}

We first note that in general we see that ProGolem favors precision over recall, while the opposite is true for FOIL and Progol. This is expected since the specific-to-general search strategy of ProGolem typically leads to more specific rules than the alternative general-to-specific strategy of FOIL and Progol. 
This can also be observed by the fact that in general, FOIL and Progol learn more clauses (rules with more disjunctions) compared to ProGolem. This leads to an increase in recall for FOIL and Progol, and the opposite for ProGolem. 

These results clearly show that the algorithms are not representation independent since the performance measures vary quite widely across different schemas.
Further, there is not an ideal representation in general since the performance measures do not vary consistently with the level of normalization. A particularly notable observation is that the runtime of all learning algorithms is severely increased when using the denormalized UW-CSE schema. Such an increase in runtime would likely discourage users who use more denormalized schemas.  

Another observation from the results is that the specific-to-general algorithm ProGolem appears to be more robust to representation change compared to the general-to-specific algorithms FOIL and Progol. For example, for UW-CSE the maximum difference in recall across representations is close to 0.3 for FOIL and Progol, but just 0.12 for ProGolem. Understanding the generality and reasons for this observation require further investigation. However, our current hypothesis is that specific-to-general algorithms are based on generalization operators that take bigger search steps and hence rely less on heuristic guidance compared to the specialization steps of FOIL and Progol. 

Finally we can see that FOIL appears to be less robust than Progol. This is likely due to the fact that FOIL is the more greedy method and is hence more sensitive to its heuristics. As an example, under the denormalized schema of UW-CSE, FOIL learns the rule $advisedby(A,B) \leftarrow course(C,B,A,D,E)$, which indicates that the student is TA for a course offered by her advisor. This rule covers 12 positive examples, so FOIL greedily adds it to the hypothesis. However, even though Progol follows the same heuristic, it is able to explore a bigger space that allows it to find better rules. 
It is also worth noting that, because of practical reasons, we restrict the search space of FOIL and Progol to 10,000 nodes. Because Progol performs a more complete search, a less restricted search space may significantly increase its performance.

%% file: exp-graphmining.tex
We have studied the representation independence 
of popular similarity search algorithms on IMDb and DBLP. 
For IMDb dataset, we create two DBs, small ($s$) and large ($l$), based on top 200 and top 5000 actors, their 1410 and 3592 associated movies produced in US from 2000 to 2012 with the rating of at least 6.0, and their 2864 and 22603 characters, respectively.
We construct IMDb graph according to the schema in Figure~\ref{fig:FreebaseIMDB}(a). 
We then transform DB to the structure in Figure~\ref{fig:FreebaseIMDB}(b). 
For DBLP, we create two DBs, small ($s$) and large ($l$), based on top 200 and top 2000 authors from 15 and 50 conferences, and their 6106 and 17116 publications from 2005-2014, respectively.
We construct the graph of DBLP based on the structure in Figure~\ref{fig:schema-dblp}(a).
We transform DBLP graph according to a schema of SIGMOD Record\footnote{\small cs.washington.edu/research/xmldatasets/} where all authors of each paper is grouped under the node {\it authors}.
We also transform DBLP graph according to the DBLP schema developed 
by L3S\footnote{\small dblp.l3s.de/d2r/}, which adds a {\it proceedings} node between 
associated {\it conference} and {\it year}, and a 
link between {\it paper} and {\it proceedings} 
if the paper appears in the proceedings.
The size of each data graph and its transformation is shown in Table~\ref{tab:graphsize}.
We randomly select 50 actors 
and 50 authors 
as queries.

\begin{table}
\centering
\scriptsize{
\begin{tabular}{c|c|rr}
DB		 & Transformation 		 & |V|	 & |E| \\
\hline
\multirow{2}{*}{IMDb$_s$} 
		 & -					 & 4474	 & 9369 \\
		 & $T_\textrm{Freebase}$ & 7681	 & 9621 \\
\hline
\multirow{2}{*}{IMDb$_l$} 
		 & -					 & 31195 & 176984 \\
		 & $T_\textrm{Freebase}$ & 61482 & 181722 \\
\hline
\multirow{3}{*}{DBLP$_s$}
		 & -					 & 8158	 & 35213 \\
		 & $T_\textrm{SIGMOD}$	 & 14264 & 41290 \\
		 & $T_\textrm{L3S}$		 & 8290	 & 41603 \\
\hline 
\multirow{3}{*}{DBLP$_l$}
		 & -					 & 19176 & 136216 \\
		 & $T_\textrm{SIGMOD}$	 & 36292 & 170354 \\
		 & $T_\textrm{L3S}$		 & 19511	& 171888 \\
\end{tabular}}
\caption{\small Size of IMDb and DBLP data graphs}
\label{tab:graphsize}
\end{table}

We compare similarity ranking between the original DBs and their transformations using RWR \cite{Tong:ICDM:06}, SimRank \cite{Jeh:KDD:02}, and PathSim \cite{Sun:VLDB:11}. Similar to RWR and SimRank, PathSim is another link-based similarity function, which captures the similarity between two entities according to the normalized path count following a specified meta-path between them, where a meta-path is a sequence of original relations that denote a new relation between the connected entities. 
Since the measurement of PathSim requires a meta-path to semantically indicate the meaning of similarity between entities in a graph, 
we pick meta-paths that represent the same or similar relations in both original and transformed DBs.
We use meta-path $AFA$ in original IMDb graph, and $ASFSA$ in the transformed graph of IMDb. 
We use $APCPA$ in original DBLP graph, $AGPCPGA$ in the graph using SIGMOD schema, and $APRCRPA$ in the graph using L3S schema, where $A$, $C$, $G$, $P$ and $R$ refer to {\em author}, {\em conference}, {\em authors}, {\em paper} and {\em proceedings}, respectively.

We use normalized Kendall's tau metric \cite{SchemaIndep:TKDE} to measure the difference between two ranked lists, which is 0 if two lists are identical and 1 if one is the reverse of the other.
Ranking methods usually return many answers 
but users focus mainly on top ranked results.
Hence, we compare the top 10 or 50 answers for each query.

\begin{table}
\centering
{\scriptsize
\begin{tabular}{r||cc|cc|cc}
 & \multicolumn{2}{c|}{$T_\textrm{Freebase}$} & \multicolumn{2}{c|}{$T_\textrm{SIGMOD}$} & \multicolumn{2}{c}{$T_\textrm{L3S}$} \\
\cline{2-7}
Algorithm & IMDb$_s$ & IMDb$_l$ & DBLP$_s$ & DBLP$_l$ & DBLP$_s$ & DBLP$_l$ \\
\hline
RWR & 		0.303 & 0.530 & 0.548 & 0.333 & 0.231 & 0.209 \\
SimRank & 	0.363 & 0.498 & 0.533 & 0.410 & 0.195 & 0.215 \\
PathSim &	0.454 & 0.557 &     {\bf 0} &     {\bf 0} & {\bf 0.060} & {\bf 0.022 }\\
\hline
RWR & 		0.211 & 0.389 & 0.527 & 0.666 & 0.359 & 0.401 \\
SimRank & 	0.435 & 0.426 & 0.526 & 0.716 & 0.389 & 0.386 \\
PathSim &	0.131 & 0.332 &     {\bf 0} &     {\bf 0} & {\bf 0.023} & {\bf 0.027} \\
\end{tabular}}
\caption{{\small Average ranking difference for top 10 (top table) and top 50 (bottom table) answers over IMDb and DBLP.}}
\label{tab:graphresults}
\end{table}

Table~\ref{tab:graphresults} shows the average ranking difference of top 10 and top 50 answers between the original and the transformed DBs of IMDb and DBLP.
Overall, none of the methods are representation independent under all transformations. 
While keeping the same information, the transformations introduced in this experiment change the topological structures of the data graphs and directly affect the computation of both RWR and SimRank.
Furthermore, since both SimRank and RWR use rather global 
information in the graph, the aggregated topological modifications
caused by the transformations largely affects their rankings. 
On the other hand, the ranking difference of PathSim is comparable to RWR and SimRank for IMDb, but almost minimal for DBLP.
This results shows that achieving representation 
independence under certain families of transformations 
is possible.
Based on the robustness of PathSim on DBLP dataset, we believe that the underlying idea of PathSim is a promising direction to attain both representation independence and effectiveness, due to the following reasons.
First, the computation of PathSim is local and constrained by a given meta-path. If transformation does not affect the part of the graph used in the computation, then the algorithm will be robust.
Second, PathSim uses meta-paths to capture 
the semantic of relationships between entities. 
By using meta-paths that represent the same type
of relationships in equivalent DBs, 
we can recover (almost) the same results over these DBs.
If a given meta-path in one DB can be transformed 
to a meta-path in another graph, PathSim 
most likely provide the same rankings over equivalent DBs.
Further, PathSim is shown to be generally more effective 
than other similarity algorithms \cite{Sun:VLDB:11}.

However, semantically equivalent meta-path 
in equivalent DBs may not exists under certain representational shifts.
For example, there is not any meta-path in the Freebase DB that has the exact semantic as meta-path $AFA$ in IMDb. 
This is because in Freebase all paths that connect {\it actor} and 
{\it film} must pass through the node {\it starring}, $S$, 
which is unique per character. But there is only one path between each actor and each film in IMDb regardless of 
the number of characters the actor has played in the film. 
This causes the ranking differences for PathSim 
over IMDb DBs.
Extending PathSim to resolve this problem is an interesting
future work.